\documentclass{moriond}

\def\NIMA{{\em Nucl. Instrum. Methods} {\bf{ A}}}

\def\PRL{\em Phys. Rev. Lett.}

\def\ra{\rightarrow}

\def\be{\begin{equation}}
\def\ee{\end{equation}}
\def\bea{\begin{eqnarray}}
\def\eea{\end{eqnarray}}

\usepackage{xspace}

\def\sPlot{\mbox{\em sPlot}\xspace}

 \def\Pb      {\ensuremath{\mathrm{b}}\xspace}                 
                  
 \def\Pd      {\ensuremath{\mathrm{d}}\xspace}

 \def\Pi      {\ensuremath{\mathrm{i}}\xspace}

 \def\Ps      {\ensuremath{\mathrm{s}}\xspace}                 
 \def\Pt      {\ensuremath{\mathrm{t}}\xspace}

\def\invfb   {\ensuremath{\mbox{\,fb}^{-1}}\xspace}

\def\dquark    {\ensuremath{\Pd}\xspace}

\def\squark    {\ensuremath{\Ps}\xspace}

\def\bquark    {\ensuremath{\Pb}\xspace}
\def\bquarkbar {\ensuremath{\overline \bquark}\xspace}

\def\tquark    {\ensuremath{\Pt}\xspace}

\def\bbbarmumuX {\ensuremath{\bquark\bquarkbar \ra \mu\mu X}\xspace}

\newcommand{\BdKpi}{\ensuremath{\Bd\to K^+\pi^-}\xspace}

\newcommand{\Bhhprime}{\ensuremath{B^0_{(s)}\to h^+h^{(')-}}\xspace}

\newcommand{\BuJpsimumuK}{\ensuremath{B^+\to J/\psi(\to \mu^+\mu^-)K^+}\xspace}

\newcommand{\BcJpsimumumunu}{\ensuremath{B^+_c\to J/\psi(\to \mu^+\mu^-)\mu^+\nu_{\mu}}\xspace}

\def\bpimumu{\ensuremath{B^{0(+)} \to \pi^{0(+)} \mu^+ \mu^-}\xspace}
\def\lbpmunu{\ensuremath{\Lambda^0_b \to p \mu^- \nu}\xspace}

\newcommand{\BsKMuNu}{\ensuremath{\ensuremath{B^0_s}\to K^- \mu^+ 
\nu_\mu}\xspace}

\newcommand{\BdPiMuNu}{\ensuremath{\ensuremath{B^0}\to \pi^- \mu^+ 
\nu_\mu}\xspace}

\newcommand{\Bmm}{\ensuremath{{B^0_{(s)}\to \mu^+\mu^-}} \xspace}
\newcommand{\Bdmm}{\ensuremath{{B^0\to \mu^+\mu^-}}\xspace}
\newcommand{\Bsmm}{\ensuremath{{B^0_{s}\to \mu^+\mu^-}}\xspace}
\newcommand{\BHsmm}{\ensuremath{{B^0_{(s),H}\to \mu^+\mu^-}}\xspace}
\newcommand{\BLsmm}{\ensuremath{{B^0_{(s),L}\to \mu^+\mu^-}}\xspace}

\newcommand{\BFBsmm}{\ensuremath{\mathcal{B}({B^0_{s}\to \mu^+\mu^-})}\xspace}
\newcommand{\BFBdmm}{\ensuremath{\mathcal{B}({B^0\to \mu^+\mu^-})}\xspace}

\newcommand{\Bd}{\ensuremath{B^0}\xspace}
\newcommand{\Bz}{\ensuremath{B^0_{(s)}}\xspace}
\newcommand{\Bs}{\ensuremath{B^0_{s}}\xspace}

\def\sqs   {\ensuremath{\protect\sqrt{s}}\xspace}
\newcommand{\tev}{\ensuremath{\mathrm{\,Te\kern -0.1em V}}\xspace}
\newcommand{\gev}{\ensuremath{\mathrm{\,Ge\kern -0.1em V}}\xspace}
\newcommand{\mev}{\ensuremath{\mathrm{\,Me\kern -0.1em V}}\xspace}
\newcommand{\kev}{\ensuremath{\mathrm{\,ke\kern -0.1em V}}\xspace}
\newcommand{\ev}{\ensuremath{\mathrm{\,e\kern -0.1em V}}\xspace}
\newcommand{\gevc}{\ensuremath{{\mathrm{\,Ge\kern -0.1em V\!/}c}}\xspace}
\newcommand{\mevc}{\ensuremath{{\mathrm{\,Me\kern -0.1em V\!/}c}}\xspace}
\newcommand{\gevcc}{\ensuremath{{\mathrm{\,Ge\kern -0.1em V\!/}c^2}}\xspace}
\newcommand{\gevgevcccc}{\ensuremath{{\mathrm{\,Ge\kern -0.1em V^2\!/}c^4}}\xspace}
\newcommand{\mevcc}{\ensuremath{{\mathrm{\,Me\kern -0.1em V\!/}c^2}}\xspace}

\def\Vtb  {\ensuremath{V_{\tquark\bquark}}\xspace}
\def\Vtq  {\ensuremath{V^{*}_{\tquark(\dquark,\squark)}}\xspace}

\def\Cppi     {\ensuremath{\mathcal{C}^{(')}_{i}}\xspace}                       % 9
\def\Oppepi   {\ensuremath{\mathcal{O}^{(')}_{i}}\xspace}                       % 9

\def\C#1      {\ensuremath{\mathcal{C}_{#1}}\xspace}                       % 9
\def\Cp#1     {\ensuremath{\mathcal{C}_{#1}^{'}}\xspace}                    % 7
\def\Cpp#1     {\ensuremath{\mathcal{C}_{#1}^{(')}}\xspace}                    % 7
\def\Opp#1     {\ensuremath{\mathcal{O}_{#1}^{(')}}\xspace}                    % 7
\def\Opp10     {\ensuremath{\mathcal{O}_{10}^{(')}}\xspace}                    % 7

\def\Ceff#1   {\ensuremath{\mathcal{C}_{#1}^{\mathrm{(eff)}}}\xspace}        % 9  
\def\Cpeff#1  {\ensuremath{\mathcal{C}_{#1}^{'\mathrm{(eff)}}}\xspace}       % 7
\def\Ope#1    {\ensuremath{\mathcal{O}_{#1}}\xspace}                       % 2
\def\Opep#1   {\ensuremath{\mathcal{O}_{#1}^{'}}\xspace}                    % 7

\def\OppS     {\ensuremath{\mathcal{O}_{S}^{(')}}\xspace}                    % 7
\def\OppP     {\ensuremath{\mathcal{O}_{P}^{(')}}\xspace}                    % 7

\def\bsll{$b \rightarrow s \ell \ell$\xspace}

\usepackage{braket}

\newcommand{\Photo}{}

\begin{document}

\vspace*{4cm}
\title{ THE BRANCHING FRACTION AND EFFECTIVE LIFETIME OF \Bmm AT LHCb WITH RUN 1 AND RUN 2 DATA}

\author{ M. MULDER \\ on behalf of the LHCb Collaboration}

\address{Nikhef, Science Park 105,\\ 1098 XG Amsterdam, Netherlands}

\maketitle\abstracts{
After Run 1 of the LHC, global fits to \bsll observables show a deviation from the Standard Model (SM) with a significance of $\sim 4$ standard devations.
An example of a \bsll process is the decay of a \Bs meson into two muons (\Bsmm).
The latest analysis of \Bmm decays by LHCb with Run 1 and Run 2 data is presented.
The \Bsmm decay is observed for the first time by a single experiment. 
In addition, the first measurement of the \Bsmm effective lifetime is performed.
No significant excess of \Bdmm decays is observed.
All results are consistent with the SM and constrain New Physics in \bsll processes. 
}

\section{Introduction}
In the SM, Flavour Changing Neutral Currents (FCNCs) occur at loop level and 
are suppressed by the GIM mechanism, and sometimes helicity suppression.
As New Physics (NP) is not necessarily suppressed, FCNCs probe physics at energies beyond the LHC centre-of-mass energy.

One such FCNC is the \bsll transition.
Several theory groups have performed global fits to \bsll observables in the Effective Field Theory (EFT) framework,
and find that the data deviate by $\sim 4$ standard deviations with respect to the SM~\cite{Altmannshofer:2014rta,Hurth:2016fbr,Descotes-Genon:2015uva}.

One example of a \bsll transition is the decay of a \Bs meson into two muons (\Bsmm). 
For \Bmm decays, the decay amplitude can be written as
\be
A(\Bmm) = \braket{\mu^+\mu^-|\mathcal{H}_{eff}|\Bz} = \frac{G_F}{\sqrt{2}} \Vtq \Vtb \sum_{i\in {10,S,P}} \Cppi  \braket{\mu^+\mu^-|\Oppepi|\Bz},
\ee
where $G_F$ is the Fermi constant, \Vtq and \Vtb are CKM elements,
 $\Cppi$ are perturbative Wilson coefficients and $\Oppepi$ are non-perturbative Wilson operators.
These operators describe axial vector (\Opp10 ), scalar (\OppS) and pseudo-scalar (\OppP) contributions respectively, 
and a prime indicates a right-handed contribution. In the SM, only \C10 contributes.
 Interestingly, contributions from \Cpp10 are helicity suppressed, while (pseudo-)scalar contributions are not. 
Therefore, \Bmm observables are very sensitive to (pseudo-)scalar NP.

As \Bmm decays are purely leptonic, the hadronic part of the amplitude reduces to the decay constant, which is well determined using lattice QCD~\cite{Aoki:2016frl}.
The branching fractions of \Bdmm and \Bsmm are consequently well predicted in the SM~\cite{Bobeth:2013uxa,Fleischer:2017ltw}:

\bea
\BFBsmm &=& (3.57 \pm 0.18) \times 10^{-9}\\
\BFBdmm &=& (1.06 \pm 0.06) \times 10^{-10}
\eea

A second observable in the \Bmm system is the effective lifetime of \Bsmm~\cite{DeBruyn:2012wk}.
\Bz mesons mix with their anti-particles and form mass (and lifetime) eigenstates.
The mass-eigenstate rate asymmetry is given by

\be
\mathcal{A}_{\Delta\Gamma}=\frac{\Gamma(\BHsmm)-\Gamma(\BLsmm)}{\Gamma(\BHsmm)+\Gamma(\BLsmm)}
\ee

In the SM, $\mathcal{A}_{\Delta\Gamma}=1$, which means that only the heavy \Bz eigenstate decays to two muons.
For \Bs mesons, the lifetime difference between their two eigenstates is measured to be 
$\Delta\Gamma/\Gamma = 0.12$~\footnote{For \Bd mesons, no lifetime difference has been observed. It is expected to be $\mathcal{O}(10^{-3})$. The effective lifetime of the \Bdmm decay is not measured due to the small lifetime difference and its small branching fraction.}
By measuring the effective lifetime from the untagged decay distribution, the mass-eigenstate rate asymmetry is probed.
The effective lifetime of \Bsmm is an independent and complementary observable to the branching fraction.

Here, the new LHCb analysis of \Bmm observables is presented~\cite{Aaij:2017vad}.
It is based on data corresponding to an integrated luminosity of 1\invfb of $pp$ collisions at a centre-of-mass energy of
\sqs = 7\tev,  2\invfb at \sqs = 8\tev and 1.4\invfb at \sqs = 13\tev.
The first two datasets are referred to as Run 1, the latter as Run 2.

In Section~\ref{sec:BF}, the branching fraction measurement is presented. 
In Section~\ref{sec:tau}, the effective lifetime measurement is presented. 
As there is a large overlap in selection and analysis strategy, 
only differences with the branching fraction measurement are mentioned in the latter Section.

\section{\Bsmm and \Bdmm branching fraction measurement}
\label{sec:BF}

\subsection{Backgrounds and selection}
\label{sec:bkg_sel}
The analysis strategy is to search for muon pairs with opposite charges,  
$m_{\mu^+\mu^-} \in [4900,6000] \mev$~\footnote{For reference, the \Bs mass is 5367 \mevcc and the \Bd mass is 5280 \mevcc.} 
and a well reconstructed, clearly displaced decay vertex.

Four background categories exist for the branching fraction measurement.
First, \bbbarmumuX decays, where both muons combine to fake a signal. This is called combinatorial background,
and all other backgrounds are hereafter referred to as peaking backgrounds.
Second, \Bhhprime decays, where $h \in [K,\pi]$ and both hadrons are misidentified as muons.  
Third, \BdPiMuNu, \BsKMuNu, and \lbpmunu, where one hadron is misidentified as a muon and the neutrino is not reconstructed. 
Fourth, \BcJpsimumumunu and \bpimumu decays, where two real muons are combined and one particle is not reconstructed. 

To reject both combinatorial and peaking backgrounds, two multivariate classifiers are used.

For the separation of \Bmm from combinatorial background, a Boosted Decision Tree (BDT) is trained.
This BDT is used to divide the sample based on signal purity.
The main discriminating variables in this BDT are based on track isolation,
 as \bbbarmumuX decays have extra tracks close to the muon tracks. 

To reject peaking backgrounds, neural networks that discriminate between muons and hadrons are trained. 
The neural network selection is optimised for the \BFBdmm measurement, as due to its lower mass it is more affected by peaking backgrounds.
 
Simulation is used for each peaking background to estimate the number of events after selection, as well as the mass and BDT shape.
The main backgrounds (\Bhhprime, \BdPiMuNu, \BsKMuNu) are also controlled using a fit to data with the $h\mu$ mass hypothesis.

\subsection{Signal calibration and normalisation}
\label{sec:sig_calib}
To convert a yield after selection into a branching fraction,
 other B decays with known branching fractions are used for normalisation. 
In addition, the mass and BDT shape of the signal have to be calibrated.

The normalisation is performed with \BuJpsimumuK and \BdKpi decays.
The two muons in the final state for the \BuJpsimumuK decay are detected similarly to the muons from \Bmm. 
The efficiency to detect the extra kaon track in the final state is corrected for using simulation.
For \BdKpi, the kinematics are very similar to \Bmm, which means the same BDT can be applied.
However, the trigger and PID selection for hadrons and muons are very different,
and are corrected for using a data-driven approach.

For the BDT calibration, fits to \BdKpi decays per BDT bin are used to determine the \Bmm BDT distributions.
The same corrections as for the normalisation are applied,
 and the BDT distribution is found to be consistent with simulation.

\begin{figure}
\begin{minipage}{0.5\linewidth}
\centerline{\includegraphics[width=1.0\linewidth]{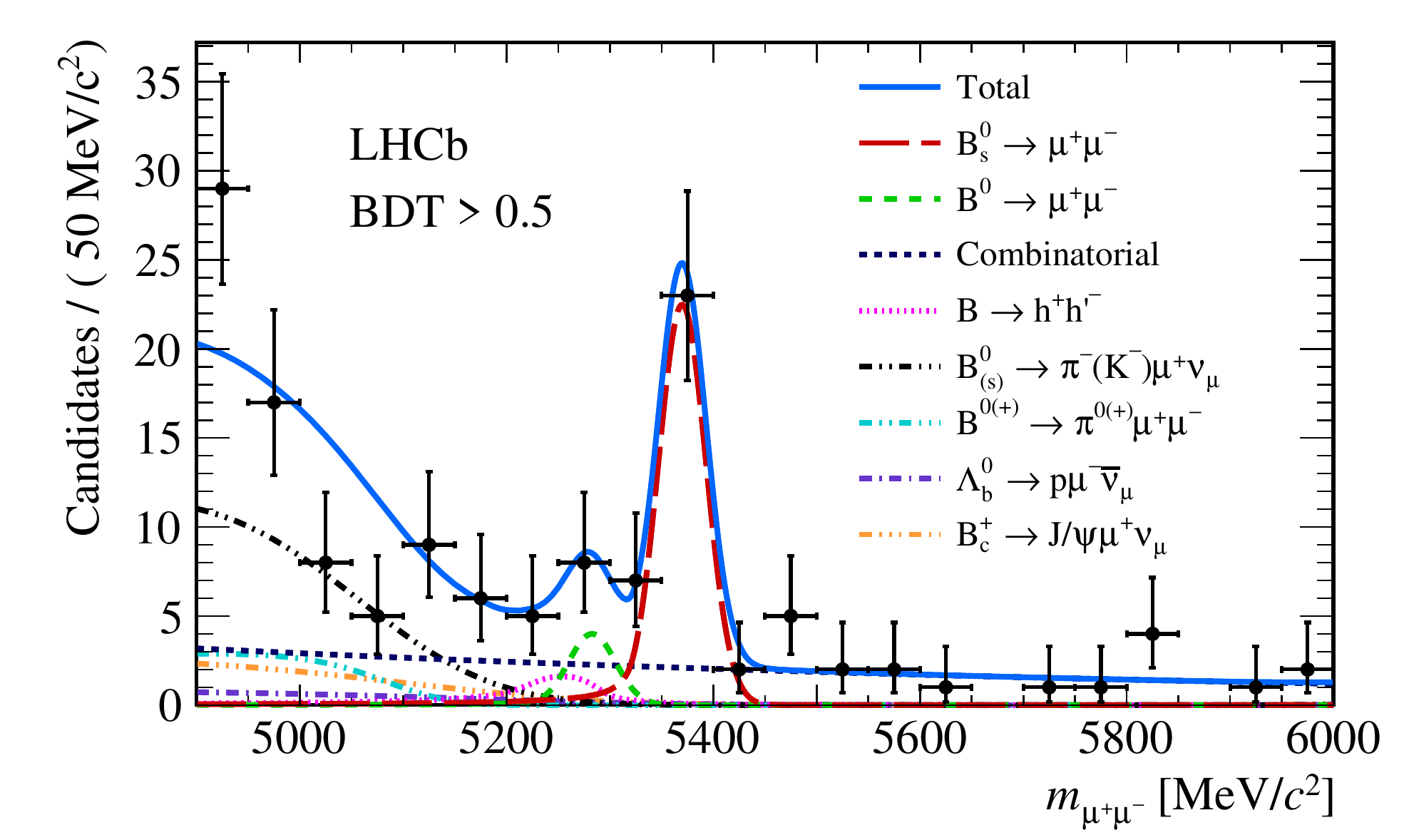}}
\end{minipage}
\begin{minipage}{0.5\linewidth}
\centerline{\includegraphics[width=1.0\linewidth]{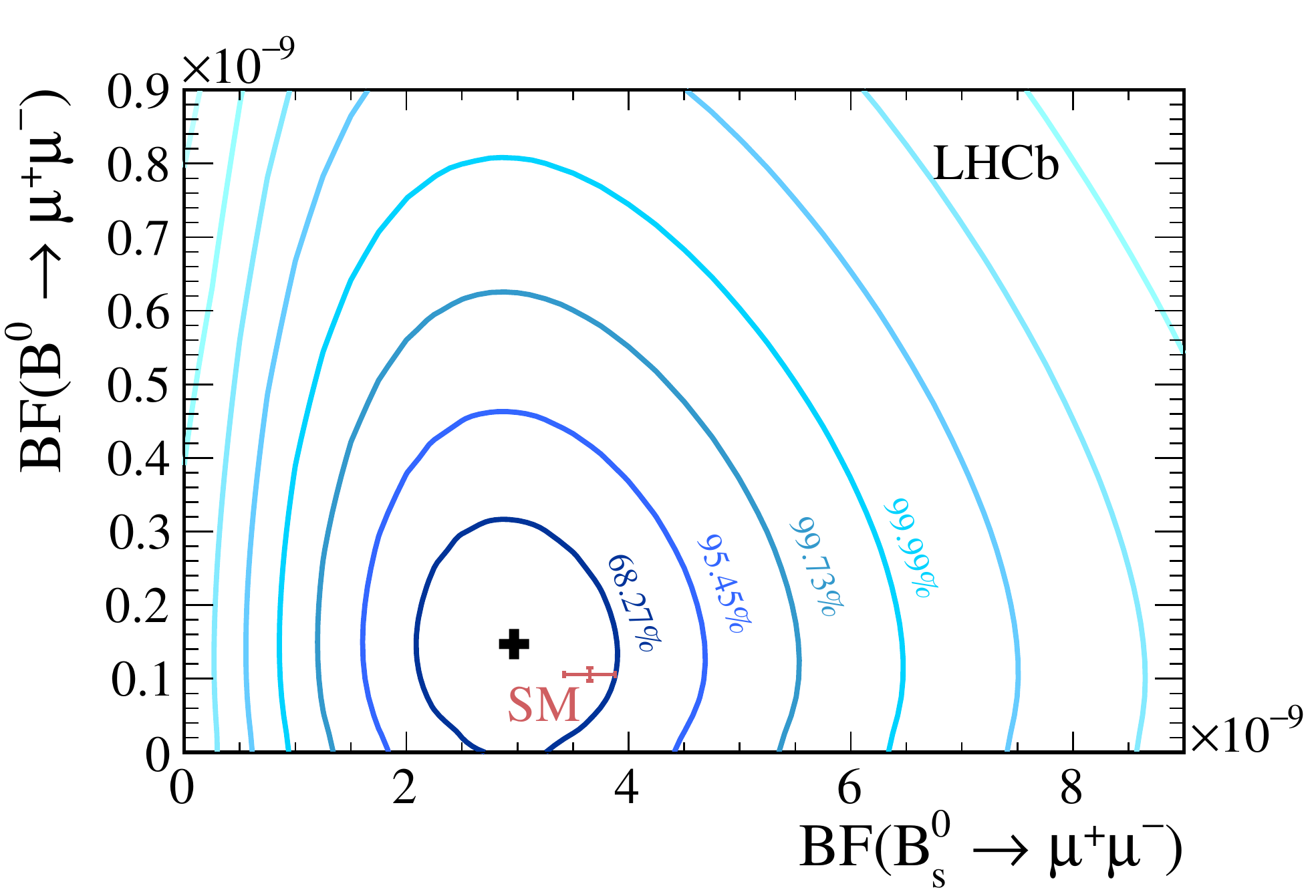}}
\end{minipage}

\caption[]{ (left) Mass distribution of selected \Bmm candidates in the four most sensitive BDT bins. The result of the fit is overlaid. (right) The 2D confidence interval in $\BFBsmm,\BFBdmm$ from this measurement. }
\label{fig:bf_fit}
\end{figure}

\subsection{Branching fraction fit and results}
To measure both branching fractions, a simultaneous maximum likelihood fit for \BFBsmm and \BFBdmm is performed to $m_{\mu^+\mu^-}$ in bins of BDT per dataset (Run 1 and Run 2).
The only free nuisance parameters are the combinatorial background shape parameters and yields.
The shape is the same for all BDT bins in a dataset, the yields are free for each BDT bin.

From the fit, the \Bsmm decay is observed with a statistical significance of 7.8 standard deviations, 
and its branching fraction is measured to be $\BFBsmm = (3.0 \pm 0.6 ^{+0.3}_{-0.2}) \times 10^{-9}$ , 
where the first uncertainty is statistical and the second systematic.
The statistical uncertainty is obtained by repeating the fit with all nuisance parameters fixed.
No significant excess of \Bdmm decays is observed, and a 95\% confidence level upper limit is determined,
$\BFBdmm < 3.4 \times 10^{-10}$.

\section{ Effective lifetime measurement of \Bsmm}
\label{sec:tau}
In contrast to the branching fraction, the effective lifetime is measured for the \Bsmm decay only.
Therefore, the selection and fitting procedure are optimised independently.

First, events with $m_{\mu^+\mu^-} < 5320 \mev$, and thus all peaking backgrounds, as well as \Bdmm events, are cut away. 
Second, the PID selection is loosened, as the mass cut makes the measurement less sensitive to peaking backgrounds.
Third, a cut is applied on the BDT, instead of using the BDT to divide the sample in bins.

With the \sPlot method~\cite{Pivk:2004ty}, a mass fit is used to background-subtract the decay time distribution.
The mass fit contains just two shapes (\Bsmm and combinatorial background).
The background-subtracted decay time distribution is fitted with an exponential, convoluted with an acceptance function. 

\begin{figure}
\begin{minipage}{0.5\linewidth}
\centerline{\includegraphics[width=1.0\linewidth]{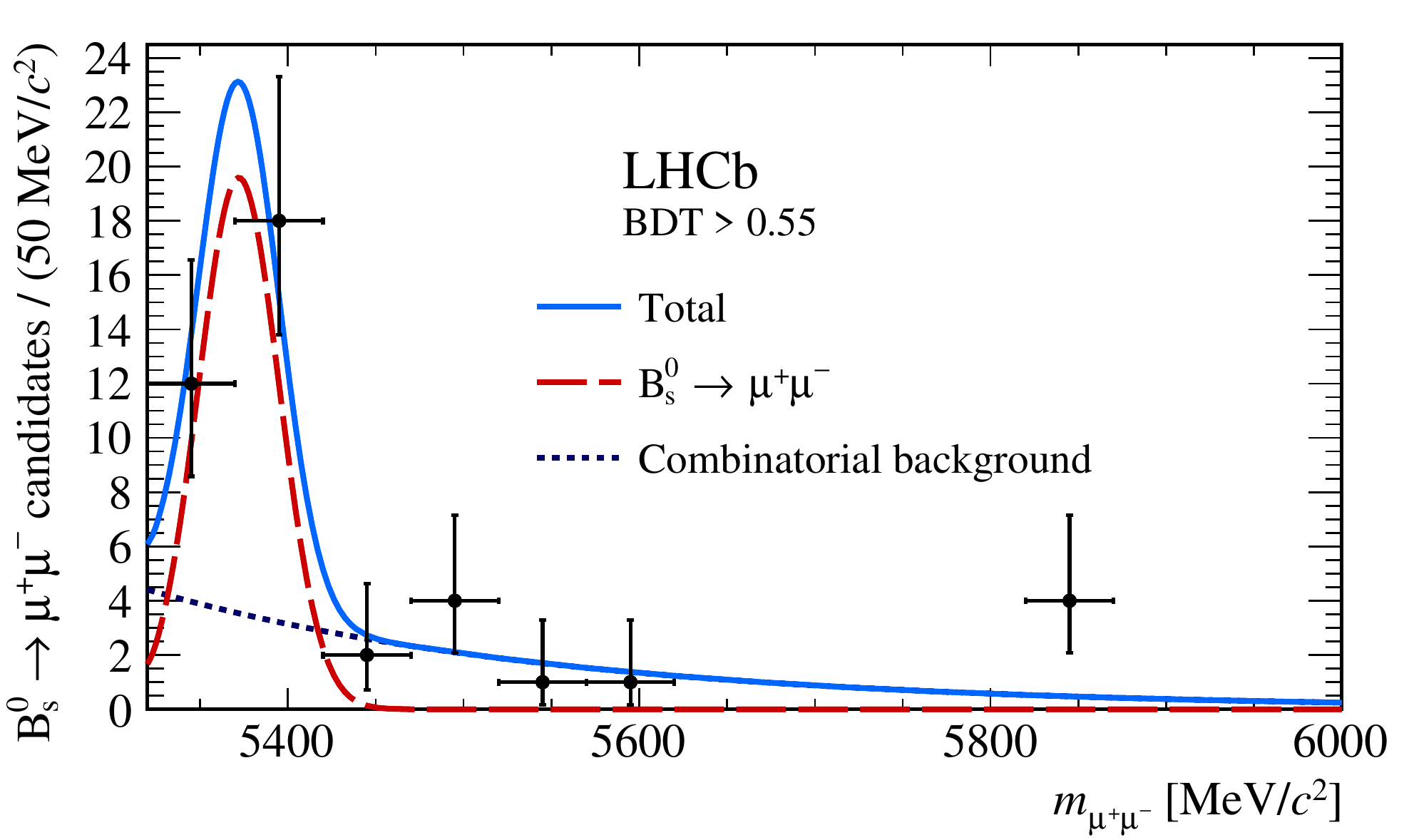}}
\end{minipage}
\begin{minipage}{0.5\linewidth}
\centerline{\includegraphics[width=1.0\linewidth]{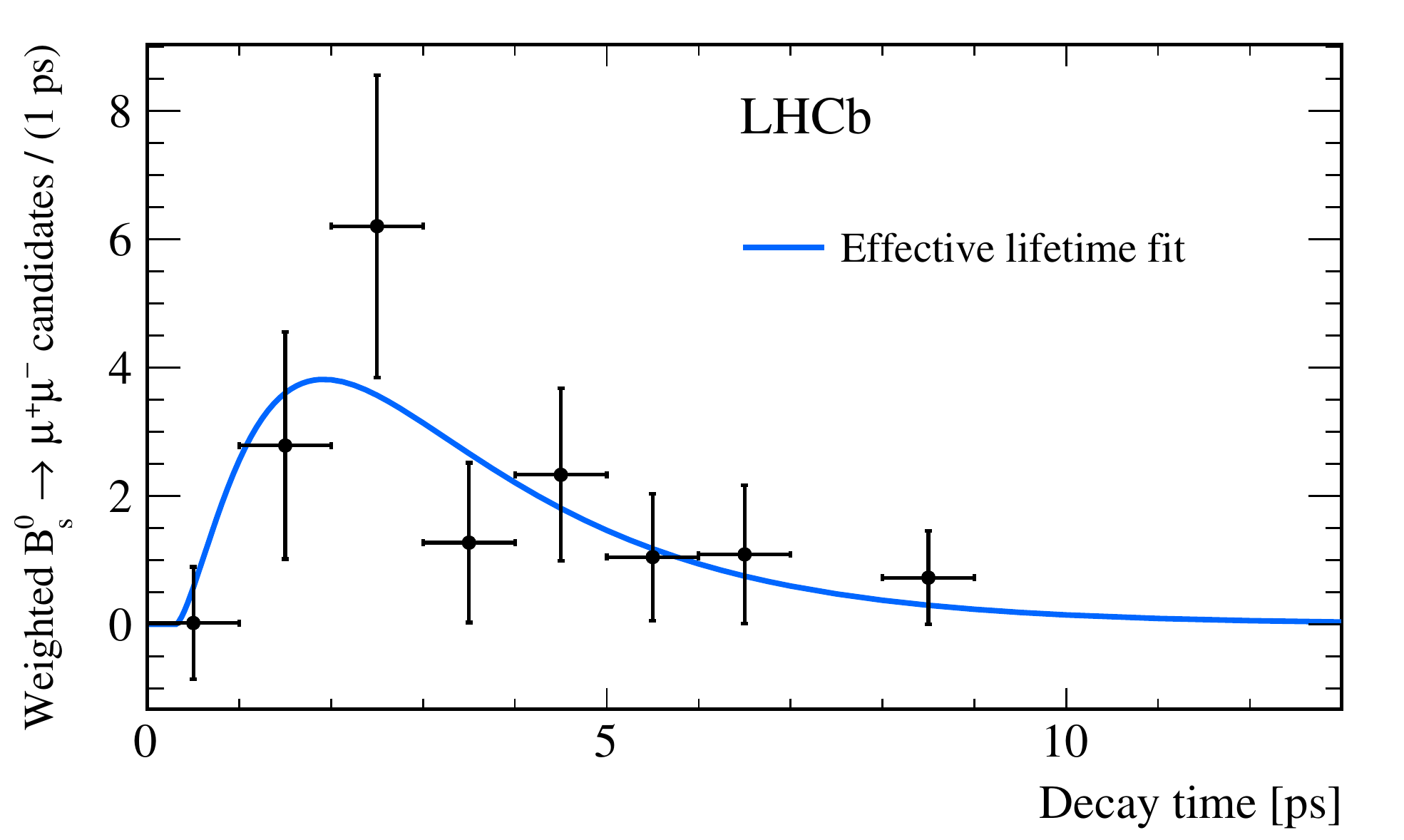}}
\end{minipage}

\caption[]{(left) Mass distribution of \Bsmm candidates. The result of the fit is overlaid. (right) Background-subtracted decay-time distribution with the fit result superimposed.}
\label{fig:tau_eff}
\end{figure}

The effective lifetime is measured for the first time and found to be $\tau (\Bsmm) = 2.04 \pm 0.44 \pm 0.05$ ps.
This measurement is consistent with the $\mathcal{A}_{\Delta\Gamma}=+1 (-1)$ hypothesis at 1.0 (1.4) standard deviations.
The current experimental precision does not set a strong constraint, 
but shows the potential LHCb will have to constrain NP using the effective lifetime.

\section{Conclusions}
The \Bsmm decay is observed with a significance of 7.8 standard deviations,
and its branching fraction is measured to be $\BFBsmm = (3.0 \pm 0.6 ^{+0.3}_{-0.2}) \times 10^{-9}$ , 
where the first uncertainty is statistical and the second systematic.
In addition, the first measurement of the \Bsmm effective lifetime is performed: 
 $\tau(\Bsmm) = 2.04 \pm 0.44 \pm 0.05$ ps.  
No significant excess of \Bdmm decays is observed, and a 95\% confidence level upper limit is determined,
$\BFBdmm < 3.4 \times 10^{-10}$.
All results are consistent with the SM and constrain New Physics in \bsll processes.
Already, these results have sparked interesting discussions~\cite{Fleischer:2017ltw,Altmannshofer:2017wqy,Bobeth:2017xry,Chiang:2017etj}.

\section*{Acknowledgments}

The author wishes to acknowledge the financial support from NWO and thank the organisers of Moriond EW 2017 for an interesting and captivating conference.

\end{document}